\documentclass[11pt,a4paper]{article}
\usepackage{jheppub} 
\usepackage{amsfonts,amssymb,latexsym,amsmath,mathrsfs,amsbsy}
\usepackage{bm}
\usepackage{slashed,bbold}
\usepackage[hhmmss]{datetime}
\usepackage{dsfont}
\usepackage{feynmf}

\usepackage[latin1]{inputenc}

\title{On Nieh-Yan Transport} 

\author[a]{Manuel Valle,}
\author[b]{Miguel \'A. V\'azquez-Mozo}

\affiliation[a]{Departamento de F\'\i sica, 
Universidad del Pa\'is Vasco UPV/EHU, \\
Apartado 644,  48080 Bilbao, Spain}
\affiliation[b]{Departamento de F\'\i sica Fundamental, Universidad de Salamanca, \\
Plaza de la Merced s/n, 37008 Salamanca, Spain}

\emailAdd{manuel.valle@ehu.es}
\emailAdd{Miguel.Vazquez-Mozo@cern.ch}

\abstract{
We study nondissipative transport 
induced by the Nieh-Yan anomaly. 
After computing the torsional terms in the 
equilibrium partition function using transgression, 
we find the constitutive relations for the covariant axial-vector,
heat, stress, and spin currents. A number of new transport effects
are found, driven by background torsion and the
spin chemical potential. Torsional constitutive relations in two-dimensional systems 
are also analyzed.
}

\arxivnumber{}

\begin{document}

\maketitle


\flushbottom


\section{Introduction}
\label{sec:intro}

Geometric torsion, a common ingredient of speculative scenarios in high energy physics~\cite{Hehl:1976kj,Shapiro:2001rz}, 
has important physical applications in the effective description of lattice dislocations in solid state physics~\cite{Katanaev:1992kh} (see~\cite{Kondo1952,Bilby:1955} for some early proposals, and~\cite{Hehl:2007bn} 
for a review). The basic 
underlying idea is to regard the lattice as a discretization of continuous space, such
that  at long distances the
lattice vectors at each point span a continuous vielbein~$e^{a}$. Lattice dislocations 
cause charge carriers circulated around then to undergo spatial translations. This phenomenon is modeled in 
the continuous theory by introducing
a background geometric torsion, which causes a ``parallelogram''
defined by two nonparallel vectors not to close. In differential geometry,
torsion is quantified by the two-form\footnote{In what follows, we denote Lorentz indices by~$a,b,c,\ldots$ 
Space-time indices are indicated by Greek fonts, whereas $i,j,k,\ldots$ are reserved for spatial indices.
In addition, to unclutter expressions, 
we drop the wedge product symbol~$\wedge$ 
throughout the paper.}
\begin{align}
T^{a}=de^{a}+\omega^{a}_{\,\,\,b}e^{b},
\label{eq:cartan_id1}
\end{align} 
where $\omega^{a}_{\,\,\,\,b}$ is the spin connection. Curvature is in turn encoded by
\begin{align}
R^{a}_{\,\,\,\,b}=d\omega^{a}_{\,\,\,b}+\omega^{a}_{\,\,\,c}\omega^{c}_{\,\,\,b},
\label{eq:cartan_id2}
\end{align} 
and effectively describes disclinations in the underlying lattice. An important 
technical point is that, in the presence 
of nonvanishing torsion,
the vielbein~$e^{a}$ and the spin connection~$\omega^{a}_{\,\,\,b}$ have to be regarded as independent fields.

For many purposes,~$e^{a}$ can be considered an Abelian gauge field, with the torsion two-form its corresponding 
(Lorentz covariant) field strength. An important difference, however, is that 
unlike a standard {\em bona fide} gauge field the vielbein is dimensionless. 
At the level of the action, torsion minimally couples to fundamental fermions via the 
axial-vector current~\cite{Hehl:1971qi,Hehl:1976kj,Audretsch:1981xn,Shapiro:2001rz} and contributes to the axial anomaly through the
correlation function of two axial-vector
currents at one loop which, being quadratically divergent, keeps memory of the relevant UV 
energy scale~$\Lambda$ of theory. Using functional methods, it was shown in~\cite{Chandia:1997hu}
that the axial anomaly receives a torsion-dependent contribution proportional to the so-called 
Nieh-Yan topological invariant~\cite{Nieh:1981ww,Nieh:1981xk}
\begin{align}
d\langle \star J_{5}\rangle&=a_{F}\mathcal{F}^{2}+a_{R}{\rm tr\,}R^{2}
+a_{\rm NY}\Lambda^{2}\eta_{ab}\Big(T^{a}T^{b}-e^{a}R^{b}_{\,\,\,c}e^{c}\Big),
\label{eq:anomaly_gauge+grav+NY}
\end{align}
where~$\mathcal{F}=d\mathcal{A}$ is the electromagnetic field strength, ``${\rm tr}$'' indicates the trace over
Lorentz indices, and the three coefficients~$a_{F}$,~$a_{R}$, and~$a_{\rm NY}$ 
are dimensionless quantities. In fact, 
using the two Cartan structure equations~\eqref{eq:cartan_id1} and~\eqref{eq:cartan_id2}, 
the Nieh-Yan term in~\eqref{eq:anomaly_gauge+grav+NY} can 
be written as an exact four-form
\begin{align}
\eta_{ab}\Big(T^{a}T^{b}-e^{a}R^{b}_{\,\,\,c}e^{c}\Big)
&=d\big(e_{a}T^{a}\big).
\label{eq:CNY}
\end{align}
Despite its multiple confirmations in the 
literature~\cite{Obukhov:1997pz,Soo:1998ev,Chandia:1998nu} and more formal 
analyses~\cite{Peeters:1999ks,Kimura:2007xa,Zanelli:2015pxa}, the torsional contribution  
in~\eqref{eq:anomaly_gauge+grav+NY} remains somewhat puzzling (see, for example,~\cite{Banados:2006fe}). 
The presence of the energy scale~$\Lambda$,
that is traced back to the peculiar dimensions of the vielbein as a putative gauge field, 
seems to conflict with the
topological and consequently infrared origin of anomalies. Moreover, while
in condensed matter 
scenarios there is a natural built-in
cutoff, it is not clear what the appropriate scale might be in the case of fundamental torsion.

Torsional anomalies have been widely studied in the context of solid state
and fluid physics~\cite{Hidaka:2012rj,Hughes:2012vg,Parrikar:2014usa,Valle:2015hfa,Nissinen:2019wmh,Nissinen:2019mkw,Nissinen:2019kld,
Huang:2019haq,Ferreiros:2020uda,Imaki:2020csc,Gallegos:2020otk,Laurila:2020yll,Gallegos:2021bzp,Manes:2020zdd,Yamamoto:2021gts,Hongo:2021ona,Nissinen:2021gke}.
In this paper we want to further this program by analyzing the consequences for transport phenomena of the 
Nieh-Yan contribution to the 't Hooft anomaly of the axial-vector current in four dimensions. 
Our strategy is to construct the equilibrium partition function from the torsional anomaly polynomial using
transgression~\cite{Jensen:2013kka,Manes:2018llx,Manes:2019fyw}. From it, 
we compute the different constitutive relations 
for axial-vector, stress, heat, and spin covariant current. In the case of the axial-vector current,
besides a vortical term, we find chiral separation effects sourced by the magnetic component
of the background torsion and the spin chemical potential.
As for
the constitutive relations for the heat and stress covariant currents, we find that they differ from
each other only
by a term proportional 
to the curl of the axial-vector external gauge field, and are therefore equal in the limit in which 
the spatial components of the external axial-vector gauge field vanish. 

We also analyze the torsional contributions to the constitutive relations of the various covariant currents in 
a~$(1+1)$-dimensional fluid in thermal equilibrium. In this case, the anomaly polynomial does not couple to 
the external gauge fields, so it is zero for theories with the same number of right- and left-handed fermion
species. For this reason, we analyze the case of a fluid of right-handed fermions coupled to an external gauge field. 
In particular, we find that the heat and stress currents are proportional to each other.

The remaining of the paper is organized as follows. In section~\ref{sec:nieh-yan-polynomial} we discuss 
the descent formalism for the covariant Nieh-Yan anomaly of
the axial-vector current. Section~\ref{sec:partition_function} is devoted to the computation of the 
equilibrium partition function of a four-dimensional 
fluid coupled to an external axial-vector gauge field, and in 
the presence of nonvanishing torsion. The torsional contributions to the constitutive relations for this
theory are analyzed in section~\ref{sec:currents_et_al}, while in section~\ref{sec:2dcase} the two-dimensional 
case is studied. Finally, we summarize and discuss our results
in section~\ref{sec:outlook}.

\section{Descent formalism for the Nieh-Yan anomaly}
\label{sec:nieh-yan-polynomial}

We begin by presenting a general 
study the Nieh-Yan anomaly using the differential geometry methods employed in the standard analysis
of quantum field theory anomalies~\cite{Zumino:1983ew}. The starting point is the anomaly polynomial in 
$D+2=6$ dimensions, obtained by adding the contributions of right- and left-handed fields with
a relative minus sign
\begin{align}
\mathcal{P}_{6}(\mathcal{F}_{R,L},d\mathcal{H})
&\equiv \left(-{i\over 24\pi^{2}}\mathcal{F}_{R}^{3}+{c_{H}\over 2}\mathcal{F}_{R}d\mathcal{H}\right)
-\left(-{i\over 24\pi^{2}}\mathcal{F}_{L}^{3}+{c_{H}\over 2}\mathcal{F}_{L}d\mathcal{H}\right)
 \nonumber \\[0.2cm]
&=-{i\over 24\pi^{2}}\Big(\mathcal{F}_{R}^{3}-\mathcal{F}_{L}^{3}\Big)
+{c_{H}\over 2}\Big(\mathcal{F}_{R}-\mathcal{F}_{L}\Big)d\mathcal{H}.
\label{eq:anomaly_polynomial}
\end{align}
Here $\mathcal{F}_{R,L}\equiv d\mathcal{A}_{R,L}$, with $\mathcal{A}_{R,L}$ the external gauge fields
coupling to microscopic right- and left-handed chiral fermions and~$c_{H}$ is 
a nonuniversal constant with dimensions 
of (energy)$^{2}$. The 
torsion-dependent part of the anomaly polynomial, on the other hand, is written in terms of the
torsional Chern-Simons three form~\cite{Zanelli:2015pxa}
\begin{align}
\mathcal{H}\equiv e_{a}T^{a}.
\end{align}
The torsional term in~\eqref{eq:anomaly_polynomial} is in fact the only closed
six-form that can be 
constructed from~$\mathcal{H}$ and~$\mathcal{F}_{R,L}$. A similar term was considered also
in refs.~\cite{Hughes:2012vg,Parrikar:2014usa} coupled to the vector field strength.

The right and left gauge fields~$\mathcal{A}_{R,L}$ can be written as the following combinations 
of the vector and axial-vector gauge fields~$\mathcal{V}$ and~$\mathcal{A}$ 
\begin{align}
\mathcal{A}_{R}&=\mathcal{V}+\mathcal{A}, \nonumber \\[0.2cm]
\mathcal{A}_{L}&=\mathcal{V}-\mathcal{A},
\end{align}
in terms of which the anomaly polynomial reads
\begin{align}
\mathcal{P}_{6}(\mathcal{F}_{V,A},\mathcal{H})&=-{i\over 4\pi^{2}}\left(\mathcal{F}_{A}\mathcal{F}_{V}^{2}
+{1\over 3}\mathcal{F}_{A}^{3}\right)+c_{H}\mathcal{F}_{A}d\mathcal{H},
\label{eq:anomaly_polynomial_VAH}
\end{align}
with~$\mathcal{F}_{V}=d\mathcal{V}$ and $\mathcal{F}_{A}=d\mathcal{A}$. This expression shows that torsion only couples
to the axial-vector gauge field. Another important feature of this anomaly polynomial is that, 
besides its gauge invariance under vector and axial-vector gauge transformations
\begin{align}
\mathcal{V}&\longrightarrow \mathcal{V}+d\alpha, \nonumber \\[0.2cm]
\mathcal{A}&\longrightarrow \mathcal{A}+d\beta,
\end{align}
it also remains invariant under shifts 
of the torsional Chern-Simons form by an arbitrary exact three-form
\begin{align}
\mathcal{H}\longrightarrow \mathcal{H}+d\gamma.
\label{eq:H-invariance}
\end{align}
In the hydrodynamical context to be explored in the following sections, the field~$\mathcal{H}$ 
can be regarded as an
external source coupling to a higher-order current form~$\mathcal{J}_{H}$~\cite{Gaiotto:2014kfa}.

When constructing the Chern-Simons five-form~$\omega^{0}_{5}(\mathcal{V},\mathcal{A},\mathcal{H})$
using the methods presented in~\cite{Manes:2018llx}, we have the freedom of
adding local counterterms. This we use to secure the invariance under vector gauge transformations, 
since the field~$\mathcal{V}$ will be eventually identified with the physical electromagnetic potential. 
In our case, given that
the torsional
term only depends on~$\mathcal{A}$ and is automatically invariant  
under vector gauge transformations, 
we only need to add the standard Abelian Bardeen counterterm, whose explicit expression can be obtained from 
Appendix B of ref.~\cite{Manes:2018llx}. 

This settled, we have the further choice of whether to preserve the invariance under torsional gauge 
transformations~\eqref{eq:H-invariance}. 
In fact, it is possible to construct a family of local counterterms that shifts the invariance of 
the torsional piece of the Chern-Simons form from axial-vector to torsional gauge invariance. 
Taking into account what was said in the previous paragraph, we write 
the following Chern-Simons five-form
\begin{align}
\omega^{0}_{5}(\mathcal{V},\mathcal{A},\mathcal{H})
&=-{i\over 4\pi^{2}}\mathcal{A}\left(\mathcal{F}_{V}^{2}+{1\over 3}\mathcal{F}_{A}^{2}\right)
+(1-a){c_{H}\over 2}\mathcal{F}_{A}\mathcal{H}+(1+a){c_{H}\over 2}\mathcal{A}d\mathcal{H},
\label{eq:omega05_ageneric}
\end{align}
where~$-1\leq a\leq 1$. This expression is invariant under vector gauge transformations, 
while breaking axial-vector gauge invariance. The torsion-dependent part, on the other hand,
exhibits the tension between axial-vector and torsional gauge transformations: 
by tuning the~$a$~parameter, we can shift its invariance 
from the first class of transformations~($a=-1$) to the second~($a=1$). Notice that there is no value of~$a$ for
which the torsional part of the 
Chern-Simons five-form remains invariant under both kinds of transformations. 
Incidentally, the pure-gauge part of the Chern-Simons form~\eqref{eq:omega05_ageneric} leads to  
the Abelian version of the Bardeen anomaly~\cite{Bardeen:1969md} (see also section~2.2 of ref.~\cite{Manes:2018llx}
for the relevant explicit expressions).

The nonlocal effective action is obtained then by integrating the Chern-Simons form 
given in~\eqref{eq:omega05_ageneric} 
on a five-dimensional
manifold~$\mathcal{M}_{5}$, whose boundary is identified with the physical spacetime
\begin{align}
\Gamma[\mathcal{V},\mathcal{A},\mathcal{H}]_{\rm CS}&=
\int\limits_{\mathcal{M}_{5}}\left[
-{i\over 4\pi^{2}}\mathcal{A}\left(\mathcal{F}_{V}^{2}+{1\over 3}\mathcal{F}_{A}^{2}\right)
+(1-a){c_{H}\over 2}\mathcal{F}_{A}\mathcal{H}+(1+a){c_{H}\over 2}\mathcal{A}d\mathcal{H}
\right].
\end{align}
To compute the consistent axial anomaly, we evaluate the variation 
of the action under axial-vector gauge
transformations,~$\delta_{\beta}\mathcal{A}=d\beta$
\begin{align}
\delta_{\beta}\Gamma[\mathcal{V},\mathcal{A},\mathcal{H}]_{\rm CS}=
-\int\limits_{\partial\mathcal{M}_{5}}\beta d\langle \star\mathcal{J}_{5}\rangle_{\rm cons},
\end{align}
and get the result
\begin{align}
d\langle \star\mathcal{J}_{5}\rangle_{\rm cons}
&={i\over 4\pi^{2}}\left(\mathcal{F}_{V}^{2}+{1\over 3}\mathcal{F}_{A}^{2}\right)
-(1+a){c_{H}\over 2}d\mathcal{H}.
\label{eq:cons_anomaly_J5_general}
\end{align}
As expected, the consistent anomaly becomes torsion-independent for~$a=-1$. 

To find the covariant
anomaly, we evaluate first the Bardeen-Zumino (BZ) term for the axial-vector current
\begin{align}
\langle \star\mathcal{J}_{5}\rangle_{\rm BZ}&\equiv -{\delta\Gamma_{\rm CS}\over\delta\mathcal{F}_{A}} \nonumber \\[0.2cm]
&={i\over 6\pi^{2}}\mathcal{A}\mathcal{F}_{A}-(1-a){c_{H}\over 2}\mathcal{H},
\label{eq:BZ_j5}
\end{align}
and apply the relation between the consistent and
covariant axial-vector currents
\begin{align}
\langle\star \mathcal{J}_{5}\rangle_{\rm cov}&=\langle\star\mathcal{J}_{5}\rangle_{\rm cons}
+\langle\star \mathcal{J}_{5}\rangle_{\rm BZ}.
\end{align}
Taking the exterior differential, we arrive at the covariant form of the axial anomaly
\begin{align}
d\langle\star \mathcal{J}_{5}\rangle_{\rm cov}&=
{i\over 4\pi^{2}}\Big(\mathcal{F}_{V}^{2}+\mathcal{F}_{A}^{2}\Big)
-c_{H}d\mathcal{H}.
\label{eq:cov_anomaly_J5_general}
\end{align}
As we see, the Nieh-Yan contribution to the
axial anomaly [cf.~\eqref{eq:CNY}] has a coefficient which is independent of
the parameter $a$. This serves as a good check of our calculation since, unlike its consistent counterpart, 
the covariant anomaly 
is insensitive to any local counterterms in the nonlocal effective action\footnote{An alternative way 
of finding 
the BZ current and the covariant anomaly is by calculating 
the variation of the Chern-Simons effective action 
under~$\delta_{B}\mathcal{A}=B$, using (see~\cite{Manes:2018llx})
\begin{align*}
\delta_{B}\Gamma[\mathcal{V},\mathcal{A},\mathcal{H}]_{\rm CS}=\int\limits_{\mathcal{M}_{5}}B\langle\star\mathcal{J}_{5}\rangle_{\rm bulk}
-\int\limits_{\partial\mathcal{M}_{5}}B\langle\star\mathcal{J}_{5}\rangle_{\rm BZ}.
\end{align*}
The covariant anomaly is then given by the value of~$-\langle\star\mathcal{J}_{5}\rangle_{\rm bulk}$  
at the boundary~$\partial\mathcal{M}_{5}$. This calculation highlights the fact that boundary terms in the 
effective action do contribute to the Bardeen-Zumino current but not to the covariant anomaly.}. 

A comparison of the results shown in eqs.~\eqref{eq:cons_anomaly_J5_general} and~\eqref{eq:cov_anomaly_J5_general} also highlights 
an intriguing feature of the Nieh-Yan anomaly. Although the torsional contribution to the
consistent axial anomaly can be cancelled by a local
counterterm without affecting the conservation of the vector current (taking $a=-1$), 
the covariant anomaly retains the Nieh-Yan term independently of the value of the parameter~$a$. 
Since the anomaly inflow argument~\cite{Jensen:2013kka} implies that physical transport of charge is 
governed by the covariant current~\cite{Manes:2019fyw}, torsional contributions to the
constitutive relations of physical currents are independent of any local counterterms added to the 
effective action.

This brings about another important issue concerning the overall normalization~$c_{H}$ of the 
torsional part of the anomaly polynomial~\eqref{eq:anomaly_polynomial_VAH}. In the general
theory of quantum field theory anomalies, it is known that
the descent method does not fix the overall normalization of the anomaly polynomial. This 
overall factor has to be
determined by a diagrammatic or functional calculation of the anomaly, or by applying the
Atiyah-Singer index theorem. In the case of the axial anomaly, this coefficient is universal, a fact
that is linked to its
infrared origin. Since it is determined by the residue of zero momentum pole in the expectation value
of the axial-vector current, it is therefore independent of the UV structure of the theory or 
model\footnote{The nonvanishing residue of this 
IR pole is precisely what explains the electromagnetic
decay of the neutral pion, despite the suppression implied by the 
Sutherland-Veltman theorem in the context of PCAC.}.

This is indeed not the case of the torsional contribution to the axial 
anomaly~\eqref{eq:cov_anomaly_J5_general}. The microscopic calculations of the Nieh-Yan 
anomaly~\cite{Chandia:1997hu} shows a strong sensitivity to the UV structure of the
theory, which leads to the fact that its overall normalization is not universal but rather
depends quadratically on the relevant UV scale of the theory. An immediate consequence  
is that the dimensionfull global factor of the torsional part of the anomaly polynomial
depends on the microscopic details of the corresponding model. This makes this quantity
nonuniversal in the sense of being model-dependent. In fact, it is set by the
energy scale of the physical effects associated to the background geometric torsion
(see also~\cite{Hughes:2012vg,Parrikar:2014usa} for
a related discussion in the context of a calculation of the effects of the 
Nieh-Yan anomaly in condensed matter
systems using a
Pauli-Villars regularization method).

This fact is actually connected with a basic mathematical issue that has been analyzed
in detail in~\cite{Parrikar:2014usa}.
Whereas both contributions on the right-hand side of the anomaly equation~\eqref{eq:cov_anomaly_J5_general}
are exact
\begin{align}
d\langle\star \mathcal{J}_{5}\rangle_{\rm cov}&=
{i\over 4\pi^{2}}d\Big(\mathcal{V}\mathcal{F}_{V}+\mathcal{A}\mathcal{F}_{A}\Big)
-c_{H}d\mathcal{H},
\end{align}
there is a glaring difference between the two terms.
The gauge part is not the differential of a globally well-defined form, so 
upon integrating it over the compact manifold~$\partial\mathcal{M}_{5}$
it is quantized in terms of the winding number of the corresponding gauge fields.
In the case of the torsional contribution, on the other hand, the three-form~$\mathcal{H}$ is globally
well-defined and therefore~$d\mathcal{H}$ gives zero when integrated over~$\partial\mathcal{M}_{5}$. 
This crucial fact preserves the topological character of the integrated axial anomaly, 
despite the presence of a dimensionfull nonuniversal quantity in the expression of the anomaly density.

After this long but necessary digression, 
we close our discussion in this section with the computation of the anomaly associated with the
torsional gauge transformations~\eqref{eq:H-invariance}.
The Ward identity for the consistent current is obtained from the variation of the Chern-Simons action
under~$\delta_{\gamma}\mathcal{H}=d\gamma$, with the result
\begin{align}
d\langle\star\mathcal{J}_{H}\rangle_{\rm cons}&=-(1-a){c_{H}\over 2}\mathcal{F}_{A}.
\end{align}
Thus, the current~$\mathcal{J}_{H}$ remains anomalous whenever~$a\neq 1$. 
The corresponding Bardeen-Zumino term is in turn 
given by (minus) the functional derivative of the action with respect to~$d\mathcal{H}$, leading to
\begin{align}
\langle\star\mathcal{J}_{H}\rangle_{\rm BZ}&=-(1+a){c_{H}\over 2}\mathcal{A}.
\end{align}
This gives the following expression for the covariant anomaly
\begin{align}
d\langle\star\mathcal{J}_{H}\rangle_{\rm cov}&=-c_{H}\mathcal{F}_{A},
\end{align}
which, as expected, is also independent of the parameter~$a$.
At this point it should be stressed however that, although intriguing, the
invariance under~\eqref{eq:H-invariance} seems to be in our context a purely accidental symmetry.

\section{The equilibrium partition function}
\label{sec:partition_function}

The Chern-Simons form~\eqref{eq:omega05_ageneric} is the basic ingredient in the construction of
the equilibrium partition function of a four-dimensional electron fluid coupled to vector and axial-vector gauge sources
and in the presence of torsion. Since the effects of the gauge part
have been widely studied in the literature, here we will focus our attention entirely 
on the torsional contributions. 
Furthermore, since 
physical charge transport is described by the
covariant currents~\cite{Manes:2018llx}, which are determined by the $a$-independent bulk
part of the effective action, 
we will set~$a=-1$ from now on in order 
to simplify expressions. This being so, our starting point is the torsional Chern-Simons form
\begin{align}
\omega^{0}_{5}(\mathcal{A},\mathcal{H})_{H}&=c_{H}\mathcal{F}_{A}\mathcal{H}.
\end{align}
To compute the partition function, we need to evaluate the
transgression form~\cite{Jensen:2013kka,Manes:2018llx,Manes:2019fyw} 
\begin{align}
\mathcal{T}_{5}(\mathcal{A},\mathcal{H};\widehat{\mathcal{A}},\widehat{\mathcal{H}})
=\omega^{0}_{5}(\mathcal{A},\mathcal{H})_{H}
-\omega^{0}_{5}(\widehat{\mathcal{A}},\widehat{\mathcal{H}})_{H},
\label{eq:transgression_def}
\end{align} 
where the two background configurations~$\{\mathcal{A},e^{a},\omega^{a}_{\,\,\,b}\}$ 
and~$\{\widehat{\mathcal{A}},\widehat{e}^{a},\widehat{\omega}^{a}_{\,\,\,b}\}$ are related by
\begin{align}
\mathcal{A}&=\widehat{\mathcal{A}}-\mu_{5}u , \nonumber \\[0.2cm]
e^{a}&=\widehat{e}^{a}-\chi^{a}u, \label{eq:final_field_conf_hats}\\[0.2cm]
\omega^{a}_{\,\,\,b}&=\widehat{\omega}^{a}_{\,\,\,b}-\mu^{a}_{\,\,\,b}u.
\nonumber
\end{align}
Here~$u=u_{\mu}dx^{\mu}$ is the fluid velocity one-form and 
$\mu_{5}$ is the chiral chemical potential, while $\chi^{a}$ and $\mu^{a}_{\,\,\,\,b}$ are respectively
interpreted as the stress and spin chemical potentials\footnote{This follows from the fact that
$e^{a}$ and~$\omega^{a}_{\,\,\,\,b}$ respectively couple to the stress and spin currents.}. 
Similar decompositions can be written for the gauge field strength and the torsion as
\begin{align}
\mathcal{F}_{A}&=\widehat{\mathcal{F}}_{A}-2\mu_{5}\omega+u\big(d+\mathfrak{a}\big)\mu_{5}, 
\nonumber \\[0.2cm]
T^{a}&=\widehat{T}^{a}-2\omega\chi^{a}
+u\Big[\big(\widehat{D}+\mathfrak{a}\big)\chi^{a}-\mu^{a}_{\,\,\,b}\widehat{e}^{b}\Big],
\end{align}
with ~$\mathfrak{a}=\imath_{u}u$ the fluid acceleration and~$\omega$ the vorticity two-form, which
is implicitly defined by the identity
\begin{align}
du=2\omega-u\mathfrak{a}.
\label{eqw:du=2omega-ua}
\end{align} 
In addition to this, we have introduced the covariant derivative of the stress chemical potential 
with respect to the hatted connection, 
$\widehat{D}\chi^{a}=d\chi^{a}+\widehat{\omega}^{a}_{\,\,\,b}\chi^{b}$. Corresponding
expressions can be derived also for the curvature two-form
\begin{align}
R^{a}_{\,\,\,b}&=\widehat{R}^{a}_{\,\,\,b}
-2\mu^{a}_{\,\,\,b}\omega
+u\big(\widehat{D}+\mathfrak{a}\big)\mu^{a}_{\,\,\,b},
\end{align}
with~$\widehat{D}\mu^{a}_{\,\,\,b}=d\mu^{a}_{\,\,\,b}
+\widehat{\omega}^{a}_{\,\,\,c}\mu^{c}_{\,\,\,b}-\mu^{a}_{\,\,\,c}\widehat{\omega}^{c}_{\,\,\,b}$,
as well as for the torsional three-form~$\mathcal{H}$
\begin{align}
\mathcal{H}&=\widehat{e}_{a}\Big(\widehat{T}^{a}-2\omega\chi^{a}\Big) 
-u\Big[\chi_{a}\widehat{T}^{a}
+\widehat{e}_{a}\big(\widehat{D}+\mathfrak{a}\big)\chi^{a}-2\omega\chi_{a}\chi^{a}
-\mu^{a}_{\,\,\,b}\widehat{e}_{a}\widehat{e}^{b}\Big]. 
\end{align}

To find an explicit expression 
for the transgression five-form~\eqref{eq:transgression_def}, we consider the following
single-parameter family of tetrads and connections interpolating between the field 
configuration~$\{\mathcal{A},e^{a},\omega^{a}_{\,\,\,b}\}$
and its hatted counterpart [cf.~\eqref{eq:final_field_conf_hats}]
\begin{align}
\mathcal{A}_{t}&=\widehat{\mathcal{A}}-t\mu_{5}u, \nonumber \\[0.2cm]
e^{a}_{t}&=\widehat{e}^{a}-t\chi^{a}u, \\[0.2cm]
(\omega_{t})^{a}_{\,\,\,b}&=\widehat{\omega}^{a}_{\,\,\,b}-t\mu^{a}_{\,\,\,b}u,
\nonumber
\end{align}
where~$0\leq t\leq 1$. 
Then, we apply the Ma\~nes-Stora-Zumino generalized transgression formula~\cite{Manes:1985df}
\begin{align}
\int_{\partial T}{\ell_{t}^{p}\over p!}\mathscr{Q}=\int_{T}{\ell_{t}^{p+1}\over (p+1)!}d\mathscr{Q}+(-1)^{p+q}
d\int_{T}{\ell_{t}^{p+1}\over (p+1)!}\mathscr{Q},
\label{eq:generalized_transgression}
\end{align}
where the even operator~$\ell_{t}$ acts by replacing exterior differential~$d$ by
\begin{align}
d_{t}=dt{d\over dt},
\end{align}
and in our case the integration domain is~$T=[0,1]$.
Taking~$\mathscr{Q}=\omega^{0}_{5}(\mathcal{A}_{t},\mathcal{H}_{t})_{H}$ with $p=q=0$, we arrive at
the following expression for the transgression five-form~\eqref{eq:transgression_def}
\begin{align}
\mathcal{T}_{5}(\mathcal{A},\mathcal{H};\widehat{\mathcal{A}},\widehat{\mathcal{H}})&=\int_{0}^{1}\ell_{t}\mathcal{P}_{6}(\mathcal{F}_{A,t},\mathcal{H}_{t})_{H}
+d\int_{0}^{1}\ell_{t}\omega^{0}_{5}(\mathcal{A}_{t},\mathcal{H}_{t})_{H},
\label{eq:transgression+MSZ}
\end{align}
where we have used that~$\mathcal{P}_{6}(\mathcal{F}_{A},\mathcal{H})_{H}\equiv c_{H}\mathcal{F}_{A}d\mathcal{H}
=d\omega^{0}_{5}(\mathcal{A},\mathcal{H})_{H}$.

At equilibrium, all hatted quantities are transverse to the fluid four-velocity~$u$~\cite{Jensen:2013kka},
so the relations~\eqref{eq:final_field_conf_hats} provide the electric-magnetic decompositions of 
the two connections and the vielbein with respect to this one-form. 
To make the notation more transparent, from now on  
all magnetic components will be denoted by boldface fonts. In particular, we set
\begin{align}
\widehat{\mathcal{A}}\equiv \boldsymbol{A}, \hspace*{1cm} \widehat{e}^{a}\equiv \pmb{\bm{e}}^{a}, 
\hspace*{1cm} \widehat{\omega}^{a}_{\,\,\,b}\equiv \boldsymbol{\omega}^{a}_{\,\,\,b}. 
\end{align}
This equilibrium condition leads to a number of identities for the magnetic parts of the
various quantities, which can be changed by tuning the external sources, while the
electric parts are fixed by the different chemical potentials and their gradients
\begin{align}
\mathcal{F}_{A}&=d\boldsymbol{A}-2\mu_{5}\omega+u\big(d+\mathfrak{a}\big)\mu_{5} \nonumber \\[0.2cm]
&\equiv \boldsymbol{B}_{A}+uE_{A}, \nonumber \\[0.2cm]
T^{a}&=d\pmb{\bm{e}}^{a}+\boldsymbol{\omega}^{a}_{\,\,\,\,b}\pmb{\bm{e}}^{b}-2\omega\chi^{a}
+u\Big[\big(\boldsymbol{D}+\mathfrak{a}\big)\chi^{a}-\mu^{a}_{\,\,\,b}\pmb{\bm{e}}^{b}\Big].
\label{eq:torsional_EandB} \\[0.2cm]
&\equiv \boldsymbol{B}^{a}+uE^{a}.
\nonumber
\end{align}
Here,  we denoted by~$\boldsymbol{D}$ the covariant derivative with respect to the magnetic part of
the spin connection~$\boldsymbol{\omega}^{a}_{\,\,\,b}$. 
For the curvature, we obtain the corresponding electric-magnetic decomposition
\begin{align}
R^{a}_{\,\,\,b}&=d\boldsymbol{\omega}^{a}_{\,\,\,b}+\boldsymbol{\omega}^{a}_{\,\,\,c}
\boldsymbol{\omega}^{c}_{\,\,\,b}-2\mu^{a}_{\,\,\,b}\omega
+u\big(\boldsymbol{D}+\mathfrak{a}\big)\mu^{a}_{\,\,\,b}.
\nonumber \\[0.2cm]
&\equiv \boldsymbol{B}^{a}_{\,\,\,b}+uE^{a}_{\,\,\,b},
\label{eq:curvature_E+B}
\end{align}
as well as for the three-form~$\mathcal{H}$
\begin{align}
\mathcal{H}&=\pmb{\bm{e}}_{a}\boldsymbol{B}^{a}-u\Big(\pmb{\bm{e}}_{a}E^{a}+\chi_{a}\boldsymbol{B}^{a}
\Big) \nonumber \\[0.2cm]
&\equiv \boldsymbol{\mathcal{H}}+uE_{H}.
\label{eq:H=H+uE}
\end{align}
Incidentally, the Bianchi identity for the torsion~$dT^{a}+\omega^{a}_{\,\,\,b}T^{b}=R^{a}_{\,\,\,b}e^{b}$ 
can be recast as 
\begin{align}
\boldsymbol{D}\boldsymbol{B}^{a}&=\boldsymbol{B}^{a}_{\,\,\,b}\pmb{\bm{e}}^{b}
-2\omega E^{a}, \nonumber \\[0.2cm]
\big(\boldsymbol{D}+\mathfrak{a}\big)E^{a}&=\boldsymbol{B}^{a}_{\,\,\,b}\chi^{b}
-E^{a}_{\,\,\,b}\pmb{\bm{e}}^{b}-\mu^{a}_{\,\,\,b}\boldsymbol{B}^{b},
\label{eq:bianchi_torsion}
\end{align}
where all electric parts are given by their values at equilibrium, shown 
in eqs.~\eqref{eq:torsional_EandB} and~\eqref{eq:curvature_E+B}.

A further consequence of setting the electric parts of the hatted quantities to zero
is that the second term on the right-hand side 
of eq.~\eqref{eq:transgression_def} vanishes
\begin{align}
\omega^{0}_{5}(\boldsymbol{A},\boldsymbol{\mathcal{H}})_{H}=0,
\end{align}
since it is a purely magnetic differential form of maximal rank. 
Plugging this into eq.~\eqref{eq:transgression+MSZ}, we get the following expression for
the equilibrium partition function
\begin{align}
W_{\rm eq}&\equiv\int\limits_{\mathcal{M}_{5}}\omega^{0}_{5}(\boldsymbol{A}-\mu_{5}u,\boldsymbol{\mathcal{H}}
+uE_{H})_{H} \nonumber \\[0.2cm]
&=\int\limits_{\mathcal{M}_{5}}\int_{0}^{1}\ell_{t}\mathcal{P}_{6}(\mathcal{A}_{t},\mathcal{H}_{t})_{H}
+\int\limits_{\partial\mathcal{M}_{5}}\int_{0}^{1}\ell_{t}\omega^{0}_{5}(\mathcal{A}_{t},\mathcal{H}_{t})_{H},
\end{align}
where in the last term 
we have used the Stokes theorem to write a boundary integral. 
This 
identity provides the standard decomposition of the equilibrium partition function into a gauge invariant
bulk piece and an anomalous boundary 
part,~$W_{\rm eq}=W_{\rm bulk}+W_{\rm bdy}$~\cite{Jensen:2013kka,Manes:2018llx,Manes:2019fyw}.
To find explicit expressions for both terms, 
we need to evaluate the action of~$\ell_{t}$ on
the anomaly polynomial and the Chern-Simons form. After a bit of algebra, we get 
\begin{align}
W_{\rm bulk}&=c_{H}\int\limits_{\mathcal{M}_{5}}
u\bigg[\mu_{5}\Big(\pmb{\bm{e}}_{a}\boldsymbol{B}^{a}_{\,\,\,c}\pmb{\bm{e}}^{c}
-\boldsymbol{B}_{a}\boldsymbol{B}^{a}\Big)
-\chi_{a}\boldsymbol{B}^{a}\Big(\boldsymbol{B}_{A}+3\mu_{5}\omega\Big)
+\mu_{5}\mu^{a}_{\,\,\,b}\pmb{\bm{e}}_{a}\pmb{\bm{e}}^{b}\omega
\nonumber \\[0.2cm]
&-2\mu_{5}\chi_{a}\chi^{a}\omega^{2} - \pmb{\bm{e}}_{a}E^{a}\Big(\boldsymbol{B}_{A}+\mu_{5}\omega\Big)
+\pmb{\bm{e}}_{a}\chi^{a}\omega E\bigg], \label{eq:Wbulk_Wbdy}\\[0.2cm]
W_{\rm bdy}&=-c_{H}\int\limits_{\partial\mathcal{M}_{5}}
u\mu_{5}\pmb{\bm{e}}_{a}\Big(\boldsymbol{B}^{a}+
\chi^{a}\omega\Big).
\nonumber
\end{align}
In writing the first identity, we have used that
\begin{align}
-2\int\limits_{\mathcal{M}_{5}}\pmb{\bm{e}}_{a}\chi^{a}\Big(\boldsymbol{B}_{A}+\mu_{5}\omega\Big)\omega=0,
\end{align}
since the integrand has no $u$-component. In addition, we also 
applied the Stokes theorem to shift a total derivative
term in~$W_{\rm bulk}$ 
into an integral over~$\partial\mathcal{M}_{5}$, which cancels an analogous term in the boundary piece~$W_{\rm bdy}$. 

\paragraph{Constraints from equilibrium.} In equilibrium, our system can be regarded 
as defined on a generic background static metric of the form~\cite{Banerjee:2012iz}
\begin{align}
ds^{2}&=-u\otimes u+g_{ij}dx^{i}\otimes dx^{j},
\label{eq:static_metric4D}
\end{align}
where~$u$ again is the fluid four-velocity one-form and all metric functions are independent 
of the time coordinate~$x^{0}$. 
Imposing that~$ds^{2}=e_{a}\otimes e^{a}$ with~$e^{a}=\pmb{\bm{e}}^{a}-u\chi^{a}$,
we derive the conditions
\begin{align}
\chi_{a}\chi^{a}&=-1, \nonumber \\[0.2cm]
\chi_{a}\pmb{\bm{e}}^{a}&=0, \label{eq:equilibrium_constraints4d}\\[0.2cm]
\pmb{\bm{e}}_{a}\otimes\pmb{\bm{e}}^{a}&=g_{ij}dx^{i}\otimes dx^{j}.
\nonumber
\end{align}
In fact, the first two identities fix the stress chemical potential in terms of the four-velocity 
\begin{align}
\chi_{a}e^{a}=u,
\label{eq:chiaea=u}
\end{align} 
from where we find~$\chi_{a}=u_{a}=(-1,0,0,0)$ and~$\chi^{a}=u^{a}=(1,0,0,0)$. 

The condition~\eqref{eq:chiaea=u} can be recast in a more useful form. 
Plugging it on the left-hand side of eq.~\eqref{eqw:du=2omega-ua}, and
separating the magnetic and electric components of the resulting equation, we find
\begin{align}
2\omega&=(\boldsymbol{D}\chi_{a})\pmb{\bm{e}}^{a}+\chi_{a}\boldsymbol{B}^{a}, \nonumber \\[0.2cm]
\mathfrak{a}&=-\chi_{a}E^{a}-\mu^{a}_{\,\,\,b}\chi^{b}\pmb{\bm{e}}_{a}.
\label{eq:2omega+a_ids}
\end{align}
In order to arrive at the second identity we have applied metric 
compatibility,~$\omega_{(ab)}=0$, as well as the
relation~$\chi_{a}\chi^{a}=-1$.
Although these constraints lead to some simplifications in the bulk and boundary partition
functions~\eqref{eq:Wbulk_Wbdy}, we should not be too hasty in implementing them. This should be done only
after taking the appropriate variations to compute the different currents. 

Finally, the thermal partition function is computed 
by taking the five-dimensional manifold to have 
topology~$\mathcal{M}_{5}=\mathcal{M}_{4}\times S^{1}_{\beta}$, with~$\beta$ the circle's length, and 
carring out the dimensional reduction onto the thermal cycle through the
substitution (cf.~\cite{Manes:2018llx,Manes:2019fyw})
\begin{align}
\int\limits_{\mathcal{M}_{4}\times S^{1}_{\beta}}u[\ldots]\longrightarrow 
\int\limits_{\mathcal{M}_{4}}{1\over T}[\ldots], 
\end{align}
where~$T$ is the local temperature. A similar replacement has to be carried out 
on the boundary integral onto~$\partial\mathcal{M}_{4}\times S^{1}_{\beta}$.
At the level of the currents, on the other hand, the prescription works by replacing $u$ with $1/T$
and the four-dimensional Hodge dual on~$\partial\mathcal{M}_{4}\times S^{1}_{\beta}$ with 
its three-dimensional counterparts
on~$\partial\mathcal{M}_{4}$.

\section{Torsional constitutive relations}
\label{sec:currents_et_al}

The constitutive relations for the various covariant currents are obtained by taking variations
of the bulk partition function~\eqref{eq:Wbulk_Wbdy} with respect to the appropriate 
sources, keeping only the boundary contributions~\cite{Jensen:2013kka,Manes:2018llx,Manes:2019fyw}. 
Let us begin with the axial-vector current.
Varying~$W_{\rm bulk}$ with respecto to the axial-vector gauge field~$\boldsymbol{A}$, we find
\begin{align}
\delta W_{\rm bulk}&=-c_{H}\int\limits_{\partial\mathcal{M}_{5}}\delta\boldsymbol{A}
u\Big(\chi_{a}\boldsymbol{B}^{a}+\pmb{\bm{e}}_{a}E^{a}\Big)+\mbox{bulk terms},
\end{align}
from where we read the expression of the axial-vector covariant current
\begin{align}
\langle \star\boldsymbol{J}_{5}\rangle_{\rm cov}
&=-c_{H}u\Big(\chi_{a}\boldsymbol{B}^{a}+\pmb{\bm{e}}_{a}E^{a}\Big).
\label{eq:j5_first}
\end{align} 
At this point we should recall that in equlibrium the 
electric parts of the axial-vector field strength, the torsion, and
the curvature are fixed in terms of the chemical potentials and their gradients by the expressions 
given in eqs.~\eqref{eq:torsional_EandB} and~\eqref{eq:curvature_E+B}.
The magnetic parts, on the
other hand, are arbitrary in the sense that they are determined by the corresponding external sources.
In fact, since we are computing torsional corrections in thermodynamical equilibrium, all transport coefficients in our
expressions are nondissipative\footnote{This suggests that the dissipative constitutive relations should include
terms proportional 
to the combinations~$E_{A}-\big(d+\boldsymbol{a}\big)\mu_{5}$ and~$E^{a}-\big(\boldsymbol{D}+\boldsymbol{a}\big)\chi^{a}+\mu^{a}_{\,\,\,b}\pmb{\bm{e}}^{b}$, which vanish in equilibrium.}. 
This in particular means that it is necessary to impose 
the first equilibrium condition in eq.~\eqref{eq:2omega+a_ids} to 
express the electric component of the torsion in terms of its magnetic part, the vorticity, 
and the spin chemical potential. Indeed, combining the identity mentioned with
the expression of~$E^{a}$ given in~\eqref{eq:torsional_EandB}, we find  
the relation
\begin{align}
-u\pmb{\bm{e}}_{a}E^{a}=u\Big(2\omega&+\mu^{a}_{\,\,\,b}\pmb{\bm{e}}_{a}\pmb{\bm{e}}^{b}-\chi_{a}\boldsymbol{B}^{a}\Big),
\label{eq:constraint_on_Ea}
\end{align} 
where we have implemented the constraint~$\chi_{a}\pmb{\bm{e}}^{a}=0$.

Substituting~\eqref{eq:constraint_on_Ea} 
into eq.~\eqref{eq:j5_first}, we arrive at the following form of the covariant axial-vector current
\begin{align}
\langle \star\boldsymbol{J}_{5}\rangle_{\rm cov}&=c_{H}u\Big(2\omega
+\mu^{a}_{\,\,\,b}\pmb{\bm{e}}_{a}\pmb{\bm{e}}^{b}-2\chi_{a}\boldsymbol{B}^{a}\Big).
\label{eq:j5_final}
\end{align}
This result shows the existence of a vortical separation effect, together with transport of chiral charge 
mediated by the spin chemical potential and the magnetic part of the torsion. 
As a nontrivial check of our result for the axial-vector covariant current, we 
evaluate the corresponding anomalous Ward identity. Implementing the 
Bianchi identities~\eqref{eq:bianchi_torsion} and after some algebra, we retrieve the 
Nieh-Yan term
\begin{align}
d\langle \star\boldsymbol{J}_{5}\rangle_{\rm cov}
&=c_{H}u\Big\{2\boldsymbol{B}_{a}\Big[\big(\boldsymbol{D}+\mathfrak{a}\big)\chi^{a}
-\mu^{a}_{\,\,\,b}\pmb{\bm{e}}^{b}\Big]
+2\chi_{a}\boldsymbol{B}^{a}_{\,\,\,b}\pmb{\bm{e}}^{b}
-\big[\big(\boldsymbol{D}+\mathfrak{a}\big)\mu^{a}_{\,\,\,b}\big]
\pmb{\bm{e}}_{a}\pmb{\bm{e}}^{b}\Big\} \nonumber \\[0.2cm]
&=
c_{H}\Big(T_{a}T^{a}-e_{a}R^{a}_{\,\,\,b}e^{b}\Big).
\end{align} 
To write these expressions we have used that the electric parts of both the torsion and the 
curvature take their equilibrium values given in eqs.~\eqref{eq:torsional_EandB} 
and~\eqref{eq:curvature_E+B}, as well as implemented the equilibrium constraint~\eqref{eq:chiaea=u}.

Our next task is the calculation of the heat and stress currents. These are obtained 
by computing the boundary variation 
of the bulk partition function induced by~$\delta u$,~$\delta\chi^{a}$,
and~$\delta e^{a}$
\begin{align}
\delta W_{\rm bulk}&=\int\limits_{\partial\mathcal{M}_{5}}\left[\delta u{\delta W_{\rm bulk}\over \delta (2\omega)}
+\delta\chi^{a}{\delta W_{\rm bulk}\over \delta E^{a}}+\delta e^{a}{\delta W_{\rm bulk}\over \delta 
\boldsymbol{B}^{a}}\right]+\mbox{bulk terms} \nonumber \\[0.2cm]
&=c_{H}\int\limits_{\partial\mathcal{M}_{5}}\left\{\delta u\,u\left(-{3\over 2}\mu_{5}\chi_{a}\boldsymbol{B}^{a}
+{1\over 2}\mu_{5}\mu^{a}_{\,\,\,b}\pmb{\bm{e}}_{a}\pmb{\bm{e}}^{b}
-2\mu_{5}\chi_{a}\chi^{a}\omega-{1\over 2}\mu_{5}\pmb{\bm{e}}_{a}E^{a}+{1\over 2}\pmb{\bm{e}}_{a}\chi^{a}E\right)
\right. \nonumber \\[0.2cm]
&+\delta \chi^{a}\pmb{\bm{e}}_{a} u\Big(\boldsymbol{B}_{A}+\mu_{5}\omega\Big)
-\delta e^{a} u\Big[2\mu_{5}\boldsymbol{B}_{a}+\chi_{a}\Big(\boldsymbol{B}_{A}+3\mu_{5}\omega\Big)\Big]\bigg\}
+\mbox{bulk terms}.
\label{eq:variationW_bulk_first_exp}
\end{align}
At this point, however, we need to realize that~$\delta u$,~$\delta\chi^{a}$,
and~$\delta e^{a}$ 
are not independent. This is a consequence of 
the equilibrium 
constrain~\eqref{eq:chiaea=u}, which implies $u\chi_{a}e^{a}=0$ and gives the following relation 
between the three variations
\begin{align}
\delta u u-\delta\chi^{a} e_{a}u-\delta e^{a}\chi_{a}u=0.
\label{eq:constraint_variations}
\end{align}
Eliminating~$\delta\chi^{a}\pmb{\bm{e}}_{a}u
=\delta\chi^{a}e_{a}u$ in favor of~$\delta u$ and~$\delta e^{a}$, we rewrite
eq.~\eqref{eq:variationW_bulk_first_exp}
as
\begin{align}
\delta W_{\rm bulk}
&=c_{H}\int\limits_{\partial\mathcal{M}_{5}}\left\{\delta u u\left(-{3\over 2}\mu_{5}\chi_{a}\boldsymbol{B}^{a}
+{1\over 2}\mu_{5}\mu^{a}_{\,\,\,b}\pmb{\bm{e}}_{a}\pmb{\bm{e}}^{b}
-{1\over 2}\mu_{5}\pmb{\bm{e}}_{a}E^{a}
+\boldsymbol{B}_{A}+3\mu_{5}\omega\right)
\right. \nonumber \\[0.2cm]
&-\delta e^{a} u\Big[2\mu_{5}\boldsymbol{B}_{a}+2\chi_{a}\Big(\boldsymbol{B}_{A}+2\mu_{5}\omega\Big)\Big]\bigg\}
+\mbox{bulk terms}.
\label{eq:variation_Wbulk_explicit}
\end{align}
The covariant heat current is given by the coefficient of the variation~$\delta u$
\begin{align}
\langle\star \boldsymbol{q}\rangle_{\rm cov}=c_{H}u\left(-{3\over 2}\mu_{5}\chi_{a}\boldsymbol{B}^{a}
+{1\over 2}\mu_{5}\mu^{a}_{\,\,\,b}\pmb{\bm{e}}_{a}\pmb{\bm{e}}^{b}
-{1\over 2}\mu_{5}\pmb{\bm{e}}_{a}E^{a}
+\boldsymbol{B}_{A}+3\mu_{5}\omega\right).
\end{align}
Here again we
eliminate the electric component of the torsion using the equilibrium condition given in
eq.~\eqref{eq:constraint_on_Ea}, to arrive at the more compact result
\begin{align}
\langle\star \boldsymbol{q}\rangle_{\rm cov}=c_{H}u\Big(\boldsymbol{B}_{A}+4\mu_{5}\omega
+\mu_{5}\mu^{a}_{\,\,\,b}\pmb{\bm{e}}_{a}\pmb{\bm{e}}^{b}
-2\mu_{5}\chi_{a}\boldsymbol{B}^{a}
\Big).
\end{align}
Interestingly, using the expression of the axial-vector current found in eq.~\eqref{eq:j5_final},
we can recast the covariant heat current in terms of the axial-vector current as (cf.~\cite{Kimura:2011ef})
\begin{align}
\langle\star \boldsymbol{q}\rangle_{\rm cov}=\mu_{5}\langle \star \boldsymbol{J}_{5}\rangle_{\rm cov}
+c_{H}u\Big(\boldsymbol{B}_{A}+2\mu_{5}\omega\Big).
\end{align}
In particular, taking~$\boldsymbol{A}=0$ we have~$\boldsymbol{B}_{A}=-2\mu_{5}\omega$ and the
heat current becomes proportional to the axial-vector 
current
\begin{align}
\langle \boldsymbol{q}\rangle_{\rm cov}=\mu_{5}\langle \boldsymbol{J}_{5}\rangle_{\rm cov}
\hspace*{2cm}(\boldsymbol{A}=0).
\end{align}

The coefficient of~$\delta e^{a}$ in~\eqref{eq:variation_Wbulk_explicit}, on the other hand, renders
the value of the stress current
\begin{align}
\langle \star\boldsymbol{\mathfrak{F}}_{a}\rangle_{\rm cov}
&=-2c_{H}u\Big[\mu_{5}\boldsymbol{B}_{a}+\chi_{a}\Big(\boldsymbol{B}_{A}+2\mu_{5}\omega\Big)\Big].
\label{eq:energy-momentum_current}
\end{align}
Unlike previous expressions, 
this is already written in terms of the chemical potentials and the magnetic parts of the axial-vector field
and the torsion, so no equilibrium constraint needs to be applied. 
Taking again~$\boldsymbol{A}=0$, the stress current becomes
\begin{align}
\langle \star\boldsymbol{\mathfrak{F}}_{a}\rangle_{\rm cov}=-2c_{H}\mu_{5}u\boldsymbol{B}_{a}
\hspace*{2cm}(\boldsymbol{A}=0),
\end{align} 
which vanishes in the limit of zero magnetic torsion.

To conclude the analysis in this section, we take variations in the bulk partition function
with respect to the spin connection
\begin{align}
\delta W_{\rm bulk}=c_{H}\int\limits_{\partial\mathcal{M}_{5}}\delta\boldsymbol{\omega}^{a}_{\,\,\,b}u\mu_{5}\pmb{\bm{e}}_{a}
\pmb{\bm{e}}^{b}+\mbox{bulk terms},
\end{align}
which gives the components of the covariant spin current
\begin{align}
\langle \star\boldsymbol{\mathfrak{S}}_{a}^{\,\,\,b}\rangle_{\rm cov}
&=c_{H}\mu_{5}u\pmb{\bm{e}}_{a}\pmb{\bm{e}}^{b}.
\end{align}
A first thing to notice in 
this result is that it is completely independent of the value of torsion, being fully generated 
by chiral imbalance. 
The corresponding anomalous Ward identity is written in terms of the energy-momentum 
current~\eqref{eq:energy-momentum_current} as
\begin{align}
D\langle \star\boldsymbol{\mathfrak{S}}_{a}^{\,\,\,b}\rangle_{\rm cov}+e^{[a}\langle \star\boldsymbol{\mathfrak{F}}^{b]}\rangle_{\rm cov}
=-c_{H}e^{a}e^{b}\mathcal{F}_{A},
\end{align}
with~$D$ the covariant derivative associated with the full connection~$\omega^{a}_{\,\,\,b}$.

Although all the previous currents have been computed from the boundary variation of the bulk effective
action, they can be alternatively obtained by adding to the appropriate variations of the
boundary partition function in~\eqref{eq:Wbulk_Wbdy} the corresponding BZ terms. For the 
axial-vector and the stress currents, these are given by
\begin{align}
\langle \star\boldsymbol{J}_{5}\rangle_{\rm BZ}&=-c_{H}T^{a}e_{a}, \nonumber \\[0.2cm]
\langle \star\boldsymbol{\mathfrak{F}}_{a}\rangle_{\rm BZ} &=-c_{H}\mathcal{F}_{A}e_{a}.
\end{align}

\section{The two-dimensional case}
\label{sec:2dcase}

Torsional constitutive relations in two dimensions can be computed along similar lines. An
important difference, however, is that unlike in the previous case now the
torsional four-form anomaly polynomial is proportional to~$d\mathcal{H}$ and does not couple 
at all to the gauge fields. 
As a consequence of this, the contributions from
right- and left-handed fermions cancel each other. This is the reason why 
here we will consider a single right-handed fermion, 
so the anomaly polynomial takes the form
\begin{align}
\mathcal{P}(\mathcal{F}_{R},\mathcal{H})&=-{1\over 4\pi}\mathcal{F}_{R}^{2}+{c_{H}\over 2}d\mathcal{H},
\end{align}
where~$c_{H}$ has again dimensions of~(energy)$^{2}$. 
The effective action constructed by integrating the corresponding Chern-Simons 
form~$\omega^{0}_{3}(\mathcal{A}_{R},\mathcal{H})$ over a three-dimensional manifold with boundary reads
\begin{align}
\Gamma[\mathcal{A}_{R},\mathcal{H}]_{\rm CS}
&=\int\limits_{\mathcal{M}_{3}}\left(
-{1\over 4\pi}\mathcal{A}_{R}\mathcal{F}_{R}+{c_{H}\over 2}\mathcal{H}\right).
\label{eq:CSeffac_2d}
\end{align}
Due to the peculiar form of the torsional contribution, 
the two-dimensional consistent anomaly does not include torsion-dependent 
terms 
\begin{align}
d\langle\star\mathcal{J}_{R}\rangle_{\rm cons}&={1\over 4\pi}\mathcal{F}_{R},
\end{align}
and neither does the right-handed BZ current
\begin{align}
\langle\star\mathcal{J}_{R}\rangle_{\rm BZ}&={1\over 4\pi}\mathcal{A}_{R}.
\end{align}
Thus, unlike in the four-dimensional case, 
the covariant anomaly does not pick up any torsional contributions either
\begin{align}
d\langle\star\mathcal{J}_{R}\rangle_{\rm cov}&={1\over 2\pi}\mathcal{F}_{R},
\end{align}
a fact that can understood on general grounds 
by noticing the absence of a two-form counterpart of the Nieh-Yan term.

All this notwithstanding, the torsional term in the Chern-Simons effective action~\eqref{eq:CSeffac_2d}
does have effects on transport, as it will be seen once we compute the equilibrium partition function and
derive the corresponding covariant currents. 
We repeat the analysis presented in section~\ref{sec:partition_function},
taking into account that on~$\mathcal{M}_{3}$ 
the magnetic component of the three-form~$\mathcal{H}$ in eq.~\eqref{eq:H=H+uE} is identically
zero, $\boldsymbol{\mathcal{H}}=0$, since maximal rank forms are purely electrical. 
The generic static two-dimensional metric, on the other hand, has the form
\begin{align}
ds^{2}=-u\otimes u+ \boldsymbol{v}\otimes \boldsymbol{v},
\end{align}
where~$\boldsymbol{v}$ is a spatial one-form. Consistency with the vielbein~$e^{a}$ results in the constraints
\begin{align}
\eta_{ab}\chi^{a}\chi^{b}&=-1, \nonumber \\[0.2cm]
\eta_{ab}\chi^{a}\pmb{\bm{e}}^{b}&=0, \label{eq:eq_constraints_2d}\\[0.2cm]
\eta_{ab}\pmb{\bm{e}}^{a}\otimes\pmb{\bm{e}}^{b}&=\boldsymbol{v}\otimes\boldsymbol{v}.
\nonumber
\end{align}
The first two identities lead again to the equilibrium constraint~\eqref{eq:chiaea=u}. Notice that 
in this case the constraints~\eqref{eq:eq_constraints_2d} are solved 
by~$\chi^{a}=(\pm 1,0)$ and~$\pmb{\bm{e}}^{a}=(0,\boldsymbol{v})$. 

Focusing on the torsional part of the Chern-Simons form
\begin{align}
\omega^{0}_{3}(\mathcal{H})_{H}={c_{H}\over 2}\mathcal{H},
\end{align}
and keeping in mind that in three 
dimensions~$\mathcal{H}=-u\big(\pmb{\bm{e}}_{a}E^{a}+\chi_{a}\boldsymbol{B}^{a}\big)$ [cf.~\eqref{eq:H=H+uE}], 
we find that the torsional part of the equilibrium partition function has no boundary piece
\begin{align}
W_{\rm eq}= W_{\rm bulk}=-{c_{H}\over 2}\int\limits_{\mathcal{M}_{3}}u\Big(\pmb{\bm{e}}_{a}E^{a}+\chi_{a}\boldsymbol{B}^{a}\Big).
\end{align}
The form of the bulk partition function shows that there are no torsional contributions either to
$\langle\star\boldsymbol{J}_{R}\rangle_{\rm cov}$ or
the covariant
spin current~$\langle \star\boldsymbol{\mathfrak{S}}_{a}^{\,\,\,b}\rangle_{\rm cov}$. 
As for the heat and stress currents, we take again
variations with respect to~$\chi^{a}$, $e^{a}$, 
and~$u$, picking up the resulting boundary terms. Since there are no explicit terms depending on~$du$, we find
\begin{align}
\delta W_{\rm bulk}=-{c_{H}\over 2}\int\limits_{\partial\mathcal{M}_{3}}\Big(\delta e^{a}\chi_{a} u-\delta\chi^{a} 
\pmb{\bm{e}}_{a}u\Big).
\end{align}
As in four dimensions, 
we have to take into account the constraint~\eqref{eq:constraint_variations}.
Eliminating~$\delta\chi^{a}\pmb{\bm{e}}_{a}u$, we get
\begin{align}
\delta W_{\rm bulk}={c_{H}\over 2}\int\limits_{\mathcal{M}_{3}}\Big(\delta u u
-2\delta e^{a}\chi_{a}u\Big).
\end{align}
From here we arrive at very simple expressions for both the covariant heat and stress currents
\begin{align}
\langle\star\boldsymbol{q}\rangle_{\rm cov}&={c_{H}\over 2}u, \nonumber \\[0.2cm]
\langle\star\boldsymbol{\mathfrak{F}}_{a}\rangle_{\rm cov}&=-c_{H}u\chi_{a}.
\end{align}
Moreover, using the first constraint in~\eqref{eq:eq_constraints_2d}, we 
derive a suggestive relation between the
torsional contributions to the stress and heat currents
\begin{align}
\langle\boldsymbol{\mathfrak{F}}^{a}\rangle_{\rm cov}=-2\chi^{a}\langle\boldsymbol{q}\rangle_{\rm cov},
\end{align}
where Hodge duals have been dropped on both sides. All these torsional terms in the 
constitutive relations of the various currents should be added to any other 
torsion-independent contributions, such as the ones studied in~\cite{Valle:2012em,Jensen:2012kj}.

\section{Closing remarks}
\label{sec:outlook}

In this paper we have studied the effects of the Nieh-Yan anomaly on the constitutive relations
of an electron fluid axially coupled to an external gauge field and in the presence of torsion. 
Beginning with the six-dimensional anomaly polynomial, we carried out the descent analysis to 
write the boundary (local) and bulk (nonlocal) contributions to the equilibrium partition function.
The covariant currents were then computed by varying the latter with respect to the 
appropriate external sources.  

Our results show the existence of Nieh-Yan-induced terms in the various covariant currents.  
To begin with, we pointed out the existence of
chiral separation effect driven by the magnetic part of the torsion, the vorticity, and the
spin chemical potential. In the presence of torsion, the two latter terms are independent, but
cancel each other for torsionless backgrounds. None of the terms in the torsional
constitutive relations for the axial-vector current actually depends on 
chiral imbalance.

As for the heat and stress currents, their expressions are particularly simple
in the limit in which the magnetic part of the axial-vector gauge field vanishes: they
are proportional to
the axial-vector and the magnetic torsion respectively. The spin current, on the other hand, 
is independent of torsion
and proportional to the chemical potential governing chiral imbalance, as well as to 
the Nieh-Yan energy scale~$c_{H}$. We have also analyzed the
two-dimensional case and found that the stress current is proportional to the heat current.  

Our result for the covariant axial-vector current of the~$(3+1)$-dimensional
theory in~\eqref{eq:j5_first} has some bearings on the results
of ref.~\cite{Ferreiros:2020uda}. It seems in principle incompatible with the ansatz for the current used in this 
reference, namely
\begin{align}
\langle\star\boldsymbol{J}_{5}\rangle_{\rm cov}&=c_{V}udu+c_{T}^{\parallel}u_{a}u_{b}T^{a}e^{b}
+c_{T}^{\perp}P_{ab}T^{a}e^{b}  \nonumber \\[0.2cm]
&=c_{T}^{\perp}\boldsymbol{B}_{a}\pmb{\bm{e}}^{a}
+u\Big(c_{V}du+c_{T}^{\parallel}\chi_{a}\boldsymbol{B}^{a}
-c_{T}^{\perp}\pmb{\bm{e}}_{a}E^{a}\Big),
\label{eq:FL_current_ansatz}
\end{align}
where~$P_{ab}=\eta_{ab}+u_{a}u_{b}$ is the projector onto the hypersurfaces orthogonal to the four-velocity.
In addition, in writing the second line
we have used the constraint~\eqref{eq:chiaea=u} in the form~$u_{a}=\chi_{a}$, which 
makes some terms vanish after implementing the first two identities in~\eqref{eq:equilibrium_constraints4d}.
It is clear that there are no values of the coefficients~$c_{T}^{\parallel}$ and~$c_{T}^{\perp}$
that can reproduce the expression found in eq.~\eqref{eq:j5_first}. 

This apparent problem is clarified once we remember that the presence of the four-velocity~$u$
breaks the invariance of our static background spacetime~\eqref{eq:static_metric4D} 
down to the subgroup preserving this vector. This means that 
in writing the most general form of the current, tensor structures built
from the electric and magnetic components of both the vierbein and the torsion 
have to be regarded as {\em independent}. To be more specific, we
have to take a general linear combination of the longitudinal and transverse projections of all three-form 
terms that are linear in the torsion and have the correct 
$\mathsf{T}$-parity\footnote{We recall that~$\pmb{\bm{e}}^{a}$ and $\boldsymbol{B}^{a}$
are $\mathsf{T}$-even, whereas~$u$,~$\chi^{a}$ and~$E^{a}$ are $\mathsf{T}$-odd. In addition, the current~$\boldsymbol{J}_{5}$ is 
$\mathsf{T}$-odd, which implies that 
its Hodge dual~$\star\boldsymbol{J}_{5}$ is $\mathsf{T}$-even.}:~$\pmb{\bm{e}}^{a}\boldsymbol{B}^{b}$,~$u\chi^{a}\boldsymbol{B}^{b}$, and~$u\pmb{\bm{e}}^{a}E^{b}$. 
In addition, we implement the equilibrium constraint~$u_{a}=\chi_{a}$, 
which implies~$u_{a}\pmb{\bm{e}}^{a}=0$ and~$P_{ab}\pmb{\bm{e}}^{b}=\pmb{\bm{e}}_{a}$, as well
as~$u_{a}\chi^{a}=-1$ and~$P_{ab}\chi^{b}=0$. 
The consequence is that the structures~$\boldsymbol{B}^{a}\pmb{\bm{e}}^{b}$ 
and~$u\pmb{\bm{e}}^{a}E^{b}$
are transverse, whereas~$u\chi^{a}\boldsymbol{B}^{b}$ is longitudinal.
Thus, the most general structure
of the axial-vector current has the form
\begin{align}
\langle\star\boldsymbol{J}_{5}\rangle_{\rm cov}&=
b_{TB}^{\perp}\boldsymbol{B}_{a}\pmb{\bm{e}}^{a}
+u\Big(c_{V}du+c_{TB}^{\parallel}\chi_{a}\boldsymbol{B}^{a}
+c_{TE}^{\perp}\pmb{\bm{e}}_{a}E^{a}\Big).
\label{eq:new_ansatz}
\end{align}
Alternatively, we can reach the same result for the 
structure of the covariant axial-vector current by taking into account that,
being the spin chemical potential of the same order as the
torsion, the most general expression of the current at equilibrium is a linear combination of
the four structures:~$\boldsymbol{B}^{a}\pmb{\bm{e}}_{a}$,~$udu$,~$u\chi^{a}\boldsymbol{B}_{a}$,
and~$u\mu^{a}_{\,\,\,b}\pmb{\bm{e}}_{a}\pmb{\bm{e}}^{b}$. We have again a total of four independent coefficients, 
which are linear combinations of the ones in eq.~\eqref{eq:new_ansatz}.

Comparing the ansatz~\eqref{eq:new_ansatz} with the one in~\eqref{eq:FL_current_ansatz}, we find that the reduced
symmetry does not force the identification of the coefficients~$b_{TB}^{\perp}$ 
and~$-c_{TE}^{\perp}$, as it is was the case 
there. 
This additional freedom 
is however crucial, since our explicit result for the axial-vector current~\eqref{eq:j5_first}
exhibits precisely the structure shown in~\eqref{eq:new_ansatz}, 
with the following values of the coefficients
\begin{align}
b_{TB}^{\perp}&=0, \nonumber \\[0.2cm]
c_{TB}^{\parallel}&=c_{TE}^{\perp}=-c_{H}.
\end{align}
Notice that in addition to the explicit term proportional to the vorticity in~\eqref{eq:new_ansatz},
there are similar contributions coming from~$u\chi^{a}\boldsymbol{B}_{a}$ and~$u\mu^{a}_{\,\,\,b}\pmb{\bm{e}}_{a}\pmb{\bm{e}}^{b}$,
after applying the equilibrium condition~\eqref{eq:constraint_on_Ea}.
In fact, the vortical term in~\eqref{eq:j5_final} entirely comes from implementing the
thermal equilibrium constraint~\eqref{eq:constraint_on_Ea} and it is therefore induced by the very presence of torsion (or, in other words, our results show that $c_{V}=0$).

The analysis presented here has provided us with a list of terms 
induced by the Nieh-Yan anomaly in the constitutive relations for the different currents, 
and therefore with a series of potentially new transport effects associated with
an effective background torsion. 
Notice, however, that the values of all the corresponding transport coefficients are proportional to the
global normalization of the torsional part of the anomaly polynomial~$c_{H}$ whose value, as we
discussed above, depends on the UV details of the concrete models. 
Given the relevance of torsion for the effective description of condensed matter systems, 
it would be interesting to go beyond a general analysis and
study these effects in specific models where a quantitative estimation 
of~$c_{H}$ is possible. This would allow to make precise predictions as to the possible
experimental signatures to be expected in realistic materials. These and other issues will 
be addressed elsewhere.

\acknowledgments
We thank Juan L. Ma\~nes for enlightening discussions and comments on the manuscript. Discussions
with Yago Ferreiros and Karl Landsteiner are also gratefully acknowledged. 
This work has been supported by 
Spanish Science Ministry grants PGC2018-094626-B-C21 (MCIU/AEI/FEDER, EU) and PGC2018-094626-B-C22
(MCIU/AEI/FEDER, EU), as well as by Basque Government
grant IT979-16.

\bibliographystyle{JHEP}
\bibliography{biblio_file}

\providecommand{\href}[2]{#2}\begingroup\raggedright\begin{thebibliography}{10}

\bibitem{Hehl:1976kj}
F.~Hehl, P.~Von Der~Heyde, G.~Kerlick and J.~Nester, \emph{{General Relativity
  with Spin and Torsion: Foundations and Prospects}},
  \href{https://doi.org/10.1103/RevModPhys.48.393}{\emph{Rev. Mod. Phys.}
  {\bfseries 48} (1976) 393}.

\bibitem{Shapiro:2001rz}
I.~Shapiro, \emph{{Physical aspects of the space-time torsion}},
  \href{https://doi.org/10.1016/S0370-1573(01)00030-8}{\emph{Phys. Rept.}
  {\bfseries 357} (2002) 113}
  [\href{https://arxiv.org/abs/hep-th/0103093}{{\ttfamily hep-th/0103093}}].

\bibitem{Katanaev:1992kh}
M.~O. Katanaev and I.~V. Volovich, \emph{{Theory of defects in solids and
  three-dimensional gravity}},
  \href{https://doi.org/10.1016/0003-4916(52)90040-7}{\emph{Annals Phys.}
  {\bfseries 216} (1992) 1}.

\bibitem{Kondo1952}
K.~Kondo, \emph{{On the geometrical and physical foundations of the theory of
  yielding}},  in \emph{{``Proceedings of the 2nd Japan National Congress for
  Applied Mechanics'', {\rm Tokyo}}}, pp.~41--47, 1952.

\bibitem{Bilby:1955}
B.~A. Bilby, R.~Bullough and E.~Smith, \emph{{Continuous Distributions of
  Dislocations: A New Application of the Methods of Non-Riemannian Geometry}},
  {\emph{Proc. R. Soc. Lond. A} {\bfseries 231} (1955) 263}.

\bibitem{Hehl:2007bn}
F.~W. Hehl and Y.~N. Obukhov, \emph{{Élie Cartan's torsion in geometry and in
  field theory, an essay}}, {\emph{Annales Fond. Broglie} {\bfseries 32} (2007)
  157} [\href{https://arxiv.org/abs/0711.1535}{{\ttfamily 0711.1535}}].

\bibitem{Hehl:1971qi}
F.~W. Hehl and B.~K. Datta, \emph{{Nonlinear spinor equation and asymmetric
  connection in general relativity}},
  \href{https://doi.org/10.1063/1.1665738}{\emph{J. Math. Phys.} {\bfseries 12}
  (1971) 1334}.

\bibitem{Audretsch:1981xn}
J.~Audretsch, \emph{{Dirac Electron in Space-times With Torsion: Spinor
  Propagation, Spin Precession, and Nongeodesic Orbits}},
  \href{https://doi.org/10.1103/PhysRevD.24.1470}{\emph{Phys. Rev. D}
  {\bfseries 24} (1981) 1470}.

\bibitem{Chandia:1997hu}
O.~Chand\'{\i}a and J.~Zanelli, \emph{{Topological invariants, instantons and
  chiral anomaly on spaces with torsion}},
  \href{https://doi.org/10.1103/PhysRevD.55.7580}{\emph{Phys. Rev. D}
  {\bfseries 55} (1997) 7580}
  [\href{https://arxiv.org/abs/hep-th/9702025}{{\ttfamily hep-th/9702025}}].

\bibitem{Nieh:1981ww}
H.~Nieh and M.~Yan, \emph{{An Identity in Riemann-Cartan Geometry}},
  \href{https://doi.org/10.1063/1.525379}{\emph{J. Math. Phys.} {\bfseries 23}
  (1982) 373}.

\bibitem{Nieh:1981xk}
H.~Nieh and M.~Yan, \emph{{Quantized Dirac Field in Curved Riemann-Cartan
  Background. 1. Symmetry Properties, Green's Function}},
  \href{https://doi.org/10.1016/0003-4916(82)90186-5}{\emph{Annals Phys.}
  {\bfseries 138} (1982) 237}.

\bibitem{Obukhov:1997pz}
Y.~N. Obukhov, E.~W. Mielke, J.~Budczies and F.~W. Hehl, \emph{{On the chiral
  anomaly in nonRiemannian space-times}},
  \href{https://doi.org/10.1007/BF02551525}{\emph{Found. Phys.} {\bfseries 27}
  (1997) 1221} [\href{https://arxiv.org/abs/gr-qc/9702011}{{\ttfamily
  gr-qc/9702011}}].

\bibitem{Soo:1998ev}
C.~Soo, \emph{{Adler-Bell-Jackiw anomaly, the Nieh-Yan form, and vacuum
  polarization}}, \href{https://doi.org/10.1103/PhysRevD.59.045006}{\emph{Phys.
  Rev. D} {\bfseries 59} (1999) 045006}
  [\href{https://arxiv.org/abs/hep-th/9805090}{{\ttfamily hep-th/9805090}}].

\bibitem{Chandia:1998nu}
O.~Chand\'{\i}a and J.~Zanelli, \emph{{Supersymmetric particle in a space-time
  with torsion and the index theorem}},
  \href{https://doi.org/10.1103/PhysRevD.58.045014}{\emph{Phys. Rev. D}
  {\bfseries 58} (1998) 045014}
  [\href{https://arxiv.org/abs/hep-th/9803034}{{\ttfamily hep-th/9803034}}].

\bibitem{Peeters:1999ks}
K.~Peeters and A.~Waldron, \emph{{Spinors on manifolds with boundary: APS index
  theorems with torsion}},
  \href{https://doi.org/10.1088/1126-6708/1999/02/024}{\emph{JHEP} {\bfseries
  02} (1999) 024} [\href{https://arxiv.org/abs/hep-th/9901016}{{\ttfamily
  hep-th/9901016}}].

\bibitem{Kimura:2007xa}
T.~Kimura, \emph{{Index theorems on torsional geometries}},
  \href{https://doi.org/10.1088/1126-6708/2007/08/048}{\emph{JHEP} {\bfseries
  08} (2007) 048} [\href{https://arxiv.org/abs/0704.2111}{{\ttfamily
  0704.2111}}].

\bibitem{Zanelli:2015pxa}
J.~Zanelli, \emph{{Chern-Simons Forms and Gravitation Theory}},  in
  \emph{{Modifications of Einstein's Theory of Gravity at Large Distances}},
  E.~Papantonopoulos, ed., {Springer Verlag}, 2015.

\bibitem{Banados:2006fe}
M.~Ba\~nados, O.~Mi{\v s}kovic and S.~Theisen, \emph{{Holographic currents in
  first order gravity and finite Fefferman-Graham expansions}},
  \href{https://doi.org/10.1088/1126-6708/2006/06/025}{\emph{JHEP} {\bfseries
  06} (2006) 025} [\href{https://arxiv.org/abs/hep-th/0604148}{{\ttfamily
  hep-th/0604148}}].

\bibitem{Hidaka:2012rj}
Y.~Hidaka, Y.~Hirono, T.~Kimura and Y.~Minami,
  \emph{{Viscoelastic-electromagnetism and Hall viscosity}},
  \href{https://doi.org/10.1093/ptep/pts063}{\emph{PTEP} {\bfseries 2013}
  (2013) 013A02} [\href{https://arxiv.org/abs/1206.0734}{{\ttfamily
  1206.0734}}].

\bibitem{Hughes:2012vg}
T.~L. Hughes, R.~G. Leigh and O.~Parrikar, \emph{{Torsional Anomalies, Hall
  Viscosity, and Bulk-boundary Correspondence in Topological States}},
  \href{https://doi.org/10.1103/PhysRevD.88.025040}{\emph{Phys. Rev. D}
  {\bfseries 88} (2013) 025040}
  [\href{https://arxiv.org/abs/1211.6442}{{\ttfamily 1211.6442}}].

\bibitem{Parrikar:2014usa}
O.~Parrikar, T.~L. Hughes and R.~G. Leigh, \emph{{Torsion, Parity-odd Response
  and Anomalies in Topological States}},
  \href{https://doi.org/10.1103/PhysRevD.90.105004}{\emph{Phys. Rev. D}
  {\bfseries 90} (2014) 105004}
  [\href{https://arxiv.org/abs/1407.7043}{{\ttfamily 1407.7043}}].

\bibitem{Valle:2015hfa}
M.~Valle, \emph{{Torsional response of relativistic fermions in $2+1$
  dimensions}}, \href{https://doi.org/10.1007/JHEP07(2015)006}{\emph{JHEP}
  {\bfseries 07} (2015) 006}
  [\href{https://arxiv.org/abs/1503.04020}{{\ttfamily 1503.04020}}].

\bibitem{Nissinen:2019wmh}
J.~Nissinen and G.~E. Volovik, \emph{{On thermal Nieh\textendash{}Yan anomaly
  in topological Weyl materials}},
  \href{https://doi.org/10.1134/S0021364019240020}{\emph{Pisma Zh. Eksp. Teor.
  Fiz.} {\bfseries 110} (2019) 797}
  [\href{https://arxiv.org/abs/1911.03382}{{\ttfamily 1911.03382}}].

\bibitem{Nissinen:2019mkw}
J.~Nissinen and G.~Volovik, \emph{{Thermal Nieh-Yan anomaly in Weyl
  superfluids}},
  \href{https://doi.org/10.1103/PhysRevResearch.2.033269}{\emph{Phys. Rev.
  Res.} {\bfseries 2} (2020) 033269}
  [\href{https://arxiv.org/abs/1909.08936}{{\ttfamily 1909.08936}}].

\bibitem{Nissinen:2019kld}
J.~Nissinen, \emph{{Emergent spacetime and gravitational Nieh-Yan anomaly in
  chiral $p+ip$ Weyl superfluids and superconductors}},
  \href{https://doi.org/10.1103/PhysRevLett.124.117002}{\emph{Phys. Rev. Lett.}
  {\bfseries 124} (2020) 117002}
  [\href{https://arxiv.org/abs/1909.05846}{{\ttfamily 1909.05846}}].

\bibitem{Huang:2019haq}
Z.-M. Huang, B.~Han and M.~Stone, \emph{{Nieh-Yan anomaly: Torsional Landau
  levels, central charge, and anomalous thermal Hall effect}},
  \href{https://doi.org/10.1103/PhysRevB.101.125201}{\emph{Phys. Rev. B}
  {\bfseries 101} (2020) 125201}
  [\href{https://arxiv.org/abs/1911.00174}{{\ttfamily 1911.00174}}].

\bibitem{Ferreiros:2020uda}
Y.~Ferreiros and K.~Landsteiner, \emph{{On chiral responses to geometric
  torsion}}, \href{https://doi.org/10.1016/j.physletb.2021.136419}{\emph{Phys.
  Lett. B} {\bfseries 819} (2021) 136419}
  [\href{https://arxiv.org/abs/2011.10535}{{\ttfamily 2011.10535}}].

\bibitem{Imaki:2020csc}
S.~Imaki and Z.~Qiu, \emph{{Chiral torsional effect with finite temperature,
  density and curvature}},
  \href{https://doi.org/10.1103/PhysRevD.102.016001}{\emph{Phys. Rev. D}
  {\bfseries 102} (2020) 016001}
  [\href{https://arxiv.org/abs/2004.11899}{{\ttfamily 2004.11899}}].

\bibitem{Gallegos:2020otk}
A.~D. Gallegos and U.~G\"ursoy, \emph{{Holographic spin liquids and Lovelock
  Chern-Simons gravity}},
  \href{https://doi.org/10.1007/JHEP11(2020)151}{\emph{JHEP} {\bfseries 11}
  (2020) 151} [\href{https://arxiv.org/abs/2004.05148}{{\ttfamily
  2004.05148}}].

\bibitem{Laurila:2020yll}
S.~Laurila and J.~Nissinen, \emph{{Torsional Landau levels and geometric
  anomalies in condensed matter Weyl systems}},
  \href{https://doi.org/10.1103/PhysRevB.102.235163}{\emph{Phys. Rev. B}
  {\bfseries 102} (2020) 235163}
  [\href{https://arxiv.org/abs/2007.10682}{{\ttfamily 2007.10682}}].

\bibitem{Gallegos:2021bzp}
A.~D. Gallegos, U.~G\"ursoy and A.~Yarom, \emph{{Hydrodynamics of spin
  currents}},
  \href{https://doi.org/10.21468/SciPostPhys.11.2.041}{\emph{SciPost Phys.}
  {\bfseries 11} (2021) 041}
  [\href{https://arxiv.org/abs/2101.04759}{{\ttfamily 2101.04759}}].

\bibitem{Manes:2020zdd}
J.~L. Ma\~nes, M.~Valle and M.~Á. V\'azquez-Mozo, \emph{{Chiral torsional
  effects in anomalous fluids in thermal equilibrium}},
  \href{https://doi.org/10.1007/JHEP05(2021)209}{\emph{JHEP} {\bfseries 05}
  (2021) 209} [\href{https://arxiv.org/abs/2012.08449}{{\ttfamily
  2012.08449}}].

\bibitem{Yamamoto:2021gts}
N.~Yamamoto and D.-L. Yang, \emph{{Helical magnetic effect and the chiral
  anomaly}}, \href{https://doi.org/10.1103/PhysRevD.103.125003}{\emph{Phys.
  Rev. D} {\bfseries 103} (2021) 125003}
  [\href{https://arxiv.org/abs/2103.13208}{{\ttfamily 2103.13208}}].

\bibitem{Hongo:2021ona}
M.~Hongo, X.-G. Huang, M.~Kaminski, M.~Stephanov and H.-U. Yee,
  \emph{{Relativistic spin hydrodynamics with torsion and linear response
  theory for spin relaxation}},
  \href{https://doi.org/10.1007/JHEP11(2021)150}{\emph{JHEP} {\bfseries 11}
  (2021) 150} [\href{https://arxiv.org/abs/2107.14231}{{\ttfamily
  2107.14231}}].

\bibitem{Nissinen:2021gke}
J.~Nissinen and G.~E. Volovik, \emph{{Anomalous chiral transport with vorticity
  and torsion: Cancellation of two mixed gravitational anomaly currents in
  rotating chiral $p+ip$ Weyl condensates}},
  \href{https://arxiv.org/abs/2111.08639}{{\ttfamily 2111.08639}}.

\bibitem{Jensen:2013kka}
K.~Jensen, R.~Loganayagam and A.~Yarom, \emph{{Anomaly inflow and thermal
  equilibrium}}, \href{https://doi.org/10.1007/JHEP05(2014)134}{\emph{JHEP}
  {\bfseries 05} (2014) 134} [\href{https://arxiv.org/abs/1310.7024}{{\ttfamily
  1310.7024}}].

\bibitem{Manes:2018llx}
J.~L. Ma\~nes, E.~Meg\'{\i}as, M.~Valle and M.~Á. V\'azquez-Mozo,
  \emph{{Non-Abelian Anomalous (Super)Fluids in Thermal Equilibrium from
  Differential Geometry}},
  \href{https://doi.org/10.1007/JHEP11(2018)076}{\emph{JHEP} {\bfseries 11}
  (2018) 076} [\href{https://arxiv.org/abs/1806.07647}{{\ttfamily
  1806.07647}}].

\bibitem{Manes:2019fyw}
J.~L. Mañes, E.~Megías, M.~Valle and M.~Á. Vázquez-Mozo, \emph{{Anomalous
  Currents and Constitutive Relations of a Chiral Hadronic Superfluid}},
  \href{https://doi.org/10.1007/JHEP12(2019)018}{\emph{JHEP} {\bfseries 12}
  (2019) 018} [\href{https://arxiv.org/abs/1910.04013}{{\ttfamily
  1910.04013}}].

\bibitem{Zumino:1983ew}
B.~Zumino, \emph{{Chiral Anomalies and Differential Geometry}},  in
  \emph{{``Relativity, groups and topology''}}, {Elsevier}, 1983.

\bibitem{Gaiotto:2014kfa}
D.~Gaiotto, A.~Kapustin, N.~Seiberg and B.~Willett, \emph{{Generalized Global
  Symmetries}}, \href{https://doi.org/10.1007/JHEP02(2015)172}{\emph{JHEP}
  {\bfseries 02} (2015) 172} [\href{https://arxiv.org/abs/1412.5148}{{\ttfamily
  1412.5148}}].

\bibitem{Bardeen:1969md}
W.~A. Bardeen, \emph{{Anomalous Ward identities in spinor field theories}},
  \href{https://doi.org/10.1103/PhysRev.184.1848}{\emph{Phys. Rev.} {\bfseries
  184} (1969) 1848}.

\bibitem{Manes:1985df}
J.~Ma\~nes, R.~Stora and B.~Zumino, \emph{{Algebraic Study of Chiral
  Anomalies}}, \href{https://doi.org/10.1007/BF01208825}{\emph{Commun. Math.
  Phys.} {\bfseries 102} (1985) 157}.

\bibitem{Banerjee:2012iz}
N.~Banerjee, J.~Bhattacharya, S.~Bhattacharyya, S.~Jain, S.~Minwalla and
  T.~Sharma, \emph{{Constraints on Fluid Dynamics from Equilibrium Partition
  Functions}}, \href{https://doi.org/10.1007/JHEP09(2012)046}{\emph{JHEP}
  {\bfseries 09} (2012) 046} [\href{https://arxiv.org/abs/1203.3544}{{\ttfamily
  1203.3544}}].

\bibitem{Kimura:2011ef}
T.~Kimura and T.~Nishioka, \emph{{The Chiral Heat Effect}},
  \href{https://doi.org/10.1143/PTP.127.1009}{\emph{Prog. Theor. Phys.}
  {\bfseries 127} (2012) 1009}
  [\href{https://arxiv.org/abs/1109.6331}{{\ttfamily 1109.6331}}].

\bibitem{Valle:2012em}
M.~Valle, \emph{{Hydrodynamics in 1+1 dimensions with gravitational
  anomalies}}, \href{https://doi.org/10.1007/JHEP08(2012)113}{\emph{JHEP}
  {\bfseries 08} (2012) 113} [\href{https://arxiv.org/abs/1206.1538}{{\ttfamily
  1206.1538}}].

\bibitem{Jensen:2012kj}
K.~Jensen, R.~Loganayagam and A.~Yarom, \emph{{Thermodynamics, gravitational
  anomalies and cones}},
  \href{https://doi.org/10.1007/JHEP02(2013)088}{\emph{JHEP} {\bfseries 02}
  (2013) 088} [\href{https://arxiv.org/abs/1207.5824}{{\ttfamily 1207.5824}}].

\end{thebibliography}\endgroup

\end{document}